\documentclass[runningheads]{llncs}
\usepackage[T1]{fontenc}
\usepackage{amsmath}
\usepackage{graphicx}
\usepackage{color}
\usepackage{url}

\usepackage{hyperref}
\usepackage{soul}
\usepackage{wrapfig}

\newcommand{\frameworkrowspace}{4pt}
\usepackage{booktabs}
\usepackage{xcolor}
\usepackage{pifont}
\usepackage{threeparttable}
\usepackage{array}
\newcommand{\cmark}{\textcolor{green!60!black}{\ding{51}}}
\newcommand{\xmark}{\textcolor{red!70!black}{\ding{55}}}

\begin{document}
\sloppy
\title{Evidence-Decision-Feedback: Theory-Driven Adaptive Scaffolding for LLM Agents}
\titlerunning{Evidence-Decision-Feedback}

\author{
    Clayton Cohn\orcidID{0000-0003-0856-9587} \and
    Siyuan Guo\orcidID{0009-0001-4305-9147} \and
    Surya Rayala\orcidID{0009-0005-8192-8138} \and
    Hanchen David Wang\orcidID{0000-0001-5990-5865} \and
    Naveeduddin Mohammed\orcidID{0000-0002-3706-2884} \and
    Umesh Timalsina\orcidID{0000-0002-5430-3993} \and
    Shruti Jain\orcidID{0009-0000-7853-0560} \and
    Angela Eeds \and
    Menton Deweese\orcidID{0000-0001-7361-3826} \and
    Pamela J. Osborn Popp\orcidID{0009-0007-5782-8895} \and
    Rebekah Stanton\orcidID{0009-0003-6562-8018} \and
    Shakeera Walker \and
    Meiyi Ma\orcidID{0000-0001-6916-8774} \and
    Gautam Biswas\orcidID{0000-0002-2752-3878}
}

\institute{
    Vanderbilt University, Nashville, TN, USA \\
    \email{\{clayton.a.cohn, siyuan.guo, surya.chand.rayala, hanchen.wang.1, naveeduddin.mohammed, umesh.timalsina, shruti.jain, angela.eeds, menton.deweese, pamela.popp, rebekah.stanton, shakeera.walker, meiyi.ma, gautam.biswas\}@vanderbilt.edu}
}

\authorrunning{Cohn et al.}


\maketitle              

\begin{abstract}
LLMs offer tremendous opportunities for pedagogical agents to help students construct knowledge and develop problem-solving skills, yet many of these agents operate on a ``one-size-fits-all'' basis, limiting their ability to personalize support. To address this, we introduce \emph{Evidence-Decision-Feedback} (EDF), a theoretical framework for adaptive scaffolding with LLM agents. EDF integrates elements of intelligent tutoring systems (ITS) and agentic behavior by organizing interactions around evidentiary inference, pedagogical decision-making, and adaptive feedback. We instantiate EDF through \emph{Copa}, a \textbf{Co}llaborative \textbf{P}eer \textbf{A}gent for STEM+C problem-solving. In an authentic high school classroom study, we show that EDF-guided interactions align feedback with students' demonstrated understanding and task mastery; promote scaffold fading; and support interpretable, evidence-grounded explanations without fostering overreliance.

\keywords{Adaptive Scaffolding \and Pedagogical Agents \and LLMs \and Agentic AI \and Multi-Agent Architectures}

\end{abstract}

\section{Introduction}
\label{sec:introduction}

Large language model (LLM) pedagogical agents afford dialogic support that aligns with learning theories emphasizing knowledge construction through social interaction, including Social Cognitive Theory~\cite{bandura1986} (SCT) and Social Constructivism~\cite{vygotsky1978mind}. However, recent studies have raised concerns: Kosmyna et al. (2025)~\cite{kosmyna2025your} reported 83\% of students receiving ChatGPT assistance while writing were unable to quote their own work, and Zhou et al. (2025)~\cite{zhou2025impact} found GPT-4 feedback on compiler errors benefited students during assisted tasks but not after feedback was withdrawn. While they can be helpful, these agents can also encourage overreliance~\cite{shi2025large}, thereby impeding sustained learning. As such, Borchers et al. (2025)~\cite{borchers2025can} concluded that LLM tutoring in its current form is ``unlikely to produce learning benefits rivaling known-to-be-effective ITS tutoring.''

Prior research also shows students often attempt to ``game'' learning systems by prioritizing task completion and performance metrics over deeper conceptual understanding~\cite{cohn2025exploring}. Such behaviors can create the appearance of mastery without corresponding gains in conceptual knowledge, diverging from learning-theoretic aims that emphasize sense-making and self-regulation. These concerns highlight the need for pedagogical agents that do more than optimize task performance, motivating a growing call within the AIED community for LLM-based agents to support students while remaining grounded in learning theory~\cite{stamper2024enhancing,cohn2026theory}. 

Such systems must be personalized to students' diverse needs while generating feedback interpretable to stakeholders, who may be wary of adopting systems whose reasoning is opaque~\cite{cohn2025cotal}. The AI community has recently embraced \emph{agentic} AI, where agents autonomously plan actions, reason over data, and execute tasks to achieve defined goals, often within \emph{multi-agent} architectures~\cite{stryker_agentic_ai_ibm}. These systems are well-suited to educational contexts, as agent reasoning can support adaptation and interpretability by grounding decisions in learner-specific data and linking feedback to learning theory. However, despite the proliferation of LLMs in the AIED community, there remains a lack of LLM-era frameworks that operationalize agentic systems in systematic, pedagogically meaningful ways.

There is a critical need for LLM-era frameworks that are (1) adaptive to diverse student needs, (2) effective in supporting student understanding, and (3) interpretable to stakeholders. In response, we introduce a theoretical framework for adaptive scaffolding---\emph{Evidence-Decision-Feedback} (EDF; Section~\ref{sec:edf})---that draws on foundational theories including Evidence-Centered Design (ECD)~\cite{mislevy2003brief}, Stealth Assessment~\cite{shute2011stealth}, SCT, and social constructivist perspectives (i.e., the Zone of Proximal Development (ZPD)~\cite{vygotsky1978mind}) to guide the design of pedagogically grounded LLM-based agents that support student learning in real time.

We developed a collaborative peer agent, \emph{Copa}, to instantiate EDF by mapping each EDF component to concrete elements of Copa's multi-agent architecture (Section~\ref{sec:copa}). In an authentic high school science classroom with $n=33$ dyads, we deployed Copa to support exploration and inquiry during problem solving within an open-ended learning environment (OELE), C2STEM \cite{hutchins2020c2stem}. Our findings suggest that EDF supports personalized, effective, and interpretable adaptive scaffolding without fostering overreliance on the agent.

\section{Background}
\label{sec:background}

\begin{table}[tb]
    \centering
    \scriptsize
    \setlength{\tabcolsep}{1.5pt}
    \renewcommand{\arraystretch}{0.84}
    \caption{Comparison of ITS and agent-based theoretical frameworks.}
    \label{tab:framework_comparison}
    \begin{tabular}{p{0.16\linewidth} c c c c c c c c c p{0.38\linewidth}}
        \toprule
        \textbf{Framework} & \textbf{OE} & \textbf{LM} & \textbf{AS} & \textbf{SR} & \textbf{INT} & \textbf{LLM} & \textbf{DC} & \textbf{DF} & \textbf{TT} & \textbf{Primary Limitation} \\
        \midrule
        Classic four-model ITS architecture~\cite{nkambou2010introduction} & \xmark & \cmark & \cmark & \xmark & \cmark & \xmark & \xmark & \xmark & \cmark & Designed for structured tutoring settings rather than open-ended, dialogue-rich LLM use. \\
        \addlinespace[\frameworkrowspace]
        ACT-R / model tracing~\cite{helander2014handbook} & \xmark & \cmark & \cmark & \xmark & \cmark & \xmark & \xmark & \xmark & \cmark & Best for well-defined domains and authoring detailed cognitive steps. \\
        \addlinespace[\frameworkrowspace]
        KLI~\cite{koedinger2012knowledge} & \xmark & \xmark & \cmark & \xmark & \cmark & \xmark & \xmark & \xmark & \cmark & Strong instructional guidance, less runtime tutoring decision support. \\
        \addlinespace[\frameworkrowspace]
        Constraint-Based Modeling~\cite{ohlsson2016constraint} & \xmark & \cmark & \cmark & \xmark & \cmark & \xmark & \xmark & \xmark & \cmark & Good fit for constraint-based diagnosis/feedback; less so for extended, dialogic interaction. \\
        \addlinespace[\frameworkrowspace]
        Open Learner Models~\cite{bull2013open} & \xmark & \cmark & \xmark & \cmark & \cmark & \xmark & \xmark & \xmark & \cmark & Emphasizes inspectable learner representations more than end-to-end tutoring control. \\
        \addlinespace[\frameworkrowspace]
        Social agency frameworks~\cite{moreno2005multimedia} & \xmark & \xmark & \xmark & \xmark & \xmark & \xmark & \cmark & \xmark & \cmark & Emphasizes relational, conversational support more than explicit learner-model orchestration. \\
        \addlinespace[\frameworkrowspace]
        SRL-oriented tutoring frameworks~\cite{azevedo2022lessons} & \xmark & \cmark & \cmark & \cmark & \cmark & \xmark & \xmark & \xmark & \cmark & Strong for self-regulation support, but not typically framed around LLM-mediated dialogue. \\
        \addlinespace[\frameworkrowspace]
        LLM-based tutor architectures~\cite{liu2025lpitutor} & \cmark & \xmark & \cmark & \xmark & \xmark & \cmark & \cmark & \cmark & \xmark & Often prioritize flexible interaction over explicit learner modeling and theory linkage. \\
        \midrule
        EDF (Ours) & \cmark & \cmark & \cmark & \cmark & \cmark & \cmark & \cmark & \cmark & \cmark & Connects learner modeling, learning theory, interpretability, and adaptive LLM feedback. \\
        \bottomrule
    \end{tabular}

    \vspace{2pt}
    \raggedright
    \scriptsize
    \textbf{Legend:} OE = Open-Ended; LM = Explicit Learner Model; AS = Adaptive Scaffolding; SR = Self-Regulation; INT = Interpretable/Inspectable; LLM = LLM-Ready; DC = Dialogue-Centered; DF = Dialogic Feedback (i.e., multi-turn, dynamic); TT = Theory-to-System Traceability.
\end{table}

In contrast to intelligent tutoring systems (ITS) that emphasize direct instruction and optimized learning paths, OELEs emphasize exploration- and inquiry-based problem solving, in which students construct knowledge through interaction with complex scenarios (e.g., ``learning-by-modeling'') ~\cite{biswas2016design}. Building effective agents for these environments necessitates robust learner modeling to support scaffolding that is both personalized to students and adaptive over time. This requires (1) attention to factors beyond cognition alone (e.g., self-regulation); (2) dialogic feedback grounded in social-cognitive and social-constructivist principles; and (3) that dialogue is treated as a first-class support mechanism, shaping what the agent says, when it probes, and how it responds. Additionally, agent decision-making and feedback must be interpretable to stakeholders and traceable to its theoretical foundations.

Table~\ref{tab:framework_comparison} compares our proposed framework to recent alternatives across several dimensions. ITS frameworks are designed primarily for well-structured domains, constrained solution paths, and tightly specified learner actions, but are less suited to OELEs where student progress is nonlinear, strategies vary, and evidence of understanding is distributed across evolving dialogue, behavior, and environment artifacts. These frameworks offer stronger theoretical grounding, interpretability, and learner modeling, but often lack dialogue-centered decision making and dialogic support for open-ended interaction. 

LLMs help address this gap by enabling agents to interpret richer learner evidence, engage in natural language dialogue, and generate context-sensitive feedback that adapts to students' evolving needs. However, many existing LLM-based agents underutilize these capabilities, relying on feedback limited in personalization, adaptivity, or multi-turn interaction~\cite{jurenka2024towards,malik2025scaffolding,thomas2025llm}. Simultaneously, advanced systems that adopt agentic or multi-agent architectures often provide less guidance on how scaffolding is grounded in pedagogy~\cite{li_edumas_2024,chu2025llm,dai_agent4edu_2025}.

EDF addresses this trade-off by combining the interpretability and theoretical grounding of ITSs with the dialogic flexibility and dynamism of LLMs. It contributes a theory-linked orchestration framework for adaptive scaffolding in open-ended environments, in which learner-state inference, pedagogical decision making, and feedback generation are integrated to address the following question: \emph{how can we support LLM-based adaptive scaffolding that is personalized, effective, and interpretable in open-ended learning environments?}

\section{Theoretical Framework}
\label{sec:edf}

\begin{figure}[htbp]
    \centering
    \includegraphics[width=1\linewidth]{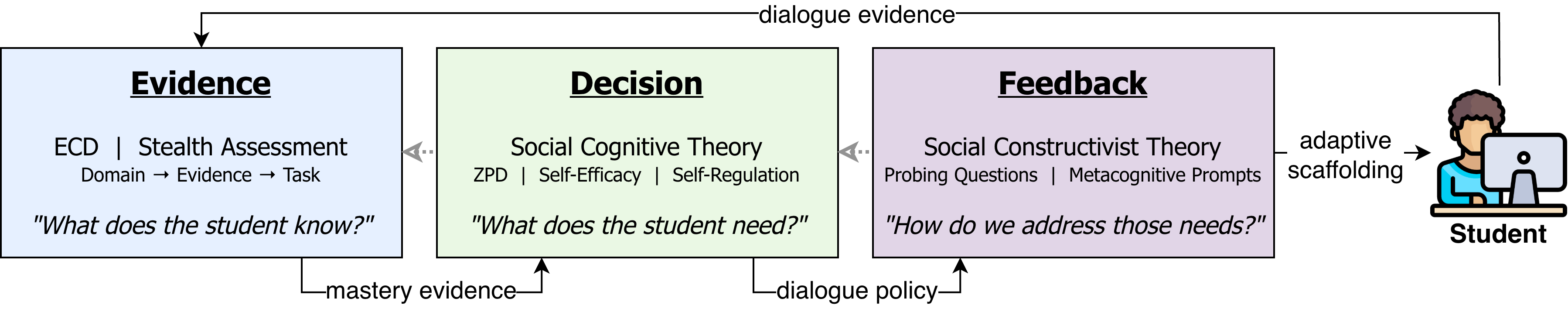}
    \caption{Evidence-Decision-Feedback (EDF).}
    \label{fig:edf}
\end{figure}

Over four years of participatory design (PD) with students, teachers, and learning scientists, we synthesized surveys, interviews, focus groups, pilot studies, and collaborative design activities into a framework aligned with established theory~\cite{cohn2024chain,cohn2024towards,cohn2025exploring,cohn2025personalizing,cohn2025multimodal,cohn2026theory,cohn2025cotal,fonteles2026jli}. We present \textbf{Evidence-Decision-Feedback (EDF)}, a principled approach to adaptive scaffolding for LLM-based pedagogical agents that addresses the limitations of existing frameworks discussed in Section~\ref{sec:background}.

EDF comprises three semi-autonomous modules that provide a blueprint for operationalizing adaptive support (shown in Figure~\ref{fig:edf}). Using ECD and Stealth Assessment, the \textbf{Evidence} module enables stakeholders to specify the student data to be monitored to continuously update a learner model. This provides \emph{mastery evidence} and is used by the \textbf{Decision} module to determine pedagogical intent (e.g., boosting self-efficacy) by generating a \emph{dialogue policy} that is consistent with SCT and within the ZPD. The \textbf{Feedback} module operationalizes this policy into adaptive scaffolding, ensuring it reflects social constructivist principles by encouraging active knowledge construction. When a student responds, the response is stored as \emph{dialogue evidence} to further update the learner model. 

Gray dotted arrows in Figure~\ref{fig:edf} denote \emph{interpretability}, a core tenet of EDF. In our context, we define interpretability as the extent to which agent output can be traced through the input-to-evidence-to-policy-to-feedback chain using observable system artifacts, thereby supporting stakeholders' ability to understand why the agent produced a given output. Using CoT prompting, we elicit the LLM's reasoning within each module so that agent feedback can be traced to dialogue policy selection, policy-to-evidentiary inferences, and evidence-to-input data. This provides a transparent account of how student input data is transformed into agent output, offering insight into agent decision-making.

EDF addresses the limitations of existing frameworks by separating learner-state inference, pedagogical decision-making, and feedback generation into discrete components, linked by a shared learner representation. This makes EDF better suited to open-ended learning, where progress is nonlinear, and problems admit multiple solutions or solution paths; and to LLM-based systems, where mappings from learner evidence to agent behavior are otherwise underspecified. Rather than prescribing fixed paths or relying on opaque end-to-end generation, EDF provides a theory-linked structure for adaptive scaffolding in which each stage remains distinct, connected, and inspectable.

\section{Copa}
\label{sec:copa}

\begin{wrapfigure}{r}{0.5\textwidth}
    \centering
    \includegraphics[width=\linewidth]{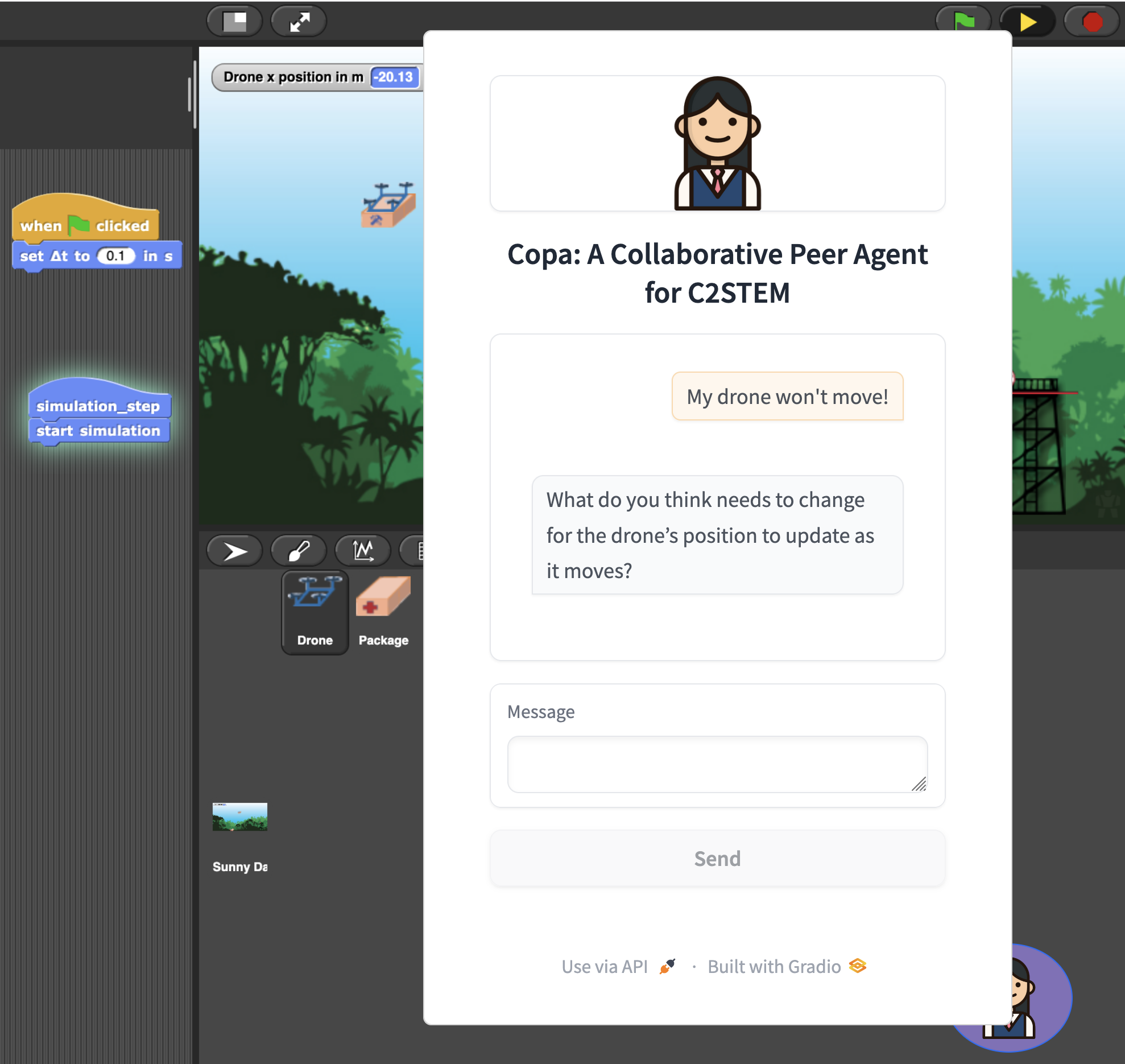}
    \caption{Copa, pictured within the C2STEM learning environment.}
    \label{fig:environment}
\end{wrapfigure}

To operationalize EDF, we developed \emph{Copa}---a \textbf{Co}llaborative \textbf{P}eer \textbf{A}gent for STEM+C learning. Copa is a \emph{knowledgeable} peer---a persona selected by students, who wanted Copa to offer both emotional support and domain competence~\cite{cohn2025exploring}. This role encourages inquiry and exploration while minimizing the dependency risk and self-efficacy loss often associated with expert tutors. Copa is agentic, using sub-agents to gather evidence, reason about learners' needs, and provide EDF-aligned scaffolding.

Copa reasons over student log and chat data, selects actions, and generates feedback while under constraints. This aligns with Jurenka et al.'s (2024)~\cite{jurenka2024towards} ``see what the student sees'' design principle and prior work suggesting \emph{bounded autonomy}~\cite{pan2025measuring}---balancing agent initiative with pedagogical guardrails---represents an ideal nexus to support learning while preserving trust~\cite{cohn2025cotal,cohn2026theory}. Copa's behavior is structured around the three EDF modules, and its multi-agent architecture is grounded in (1) participatory design, which specifies the desired agent behavior and input data; and (2) prior work demonstrating modular, multi-agent systems can improve transparency and task decomposition in educational settings~\cite{li_edumas_2024,chu2025llm,dai_agent4edu_2025}.

Copa is embedded within the C2STEM learning environment (see Figure~\ref{fig:environment}), which targets 1- and 2-D kinematics through open-ended dyadic computational modeling tasks. Students simulate the motion of objects such as a truck (1-D) and a drone (2-D) using kinematic equations and block-based code, interacting with Copa as they see fit. C2STEM generates rich log data to support agent reasoning, including \emph{logged actions} (e.g., editing code blocks), \emph{model state} (the code blocks on screen), and \emph{task context} (the model component currently being addressed, e.g., variable initialization). 

Logged actions are mapped to \emph{processed actions} to provide the LLM with a semantically meaningful representation (e.g., \texttt{set\_blk1234\_vel\_4} $\rightarrow$ \texttt{Set Velocity to 4 m/s}). Logs also allow us to track student progress by defining rubric criteria for each task. Importantly, we distinguish between students' \emph{understanding}---their verbalized knowledge to the agent---and their \emph{task mastery}---performance measured as the percentage score achieved for a given task.

\begin{figure}[t] 
    \centering 
    \includegraphics[width=1\linewidth]{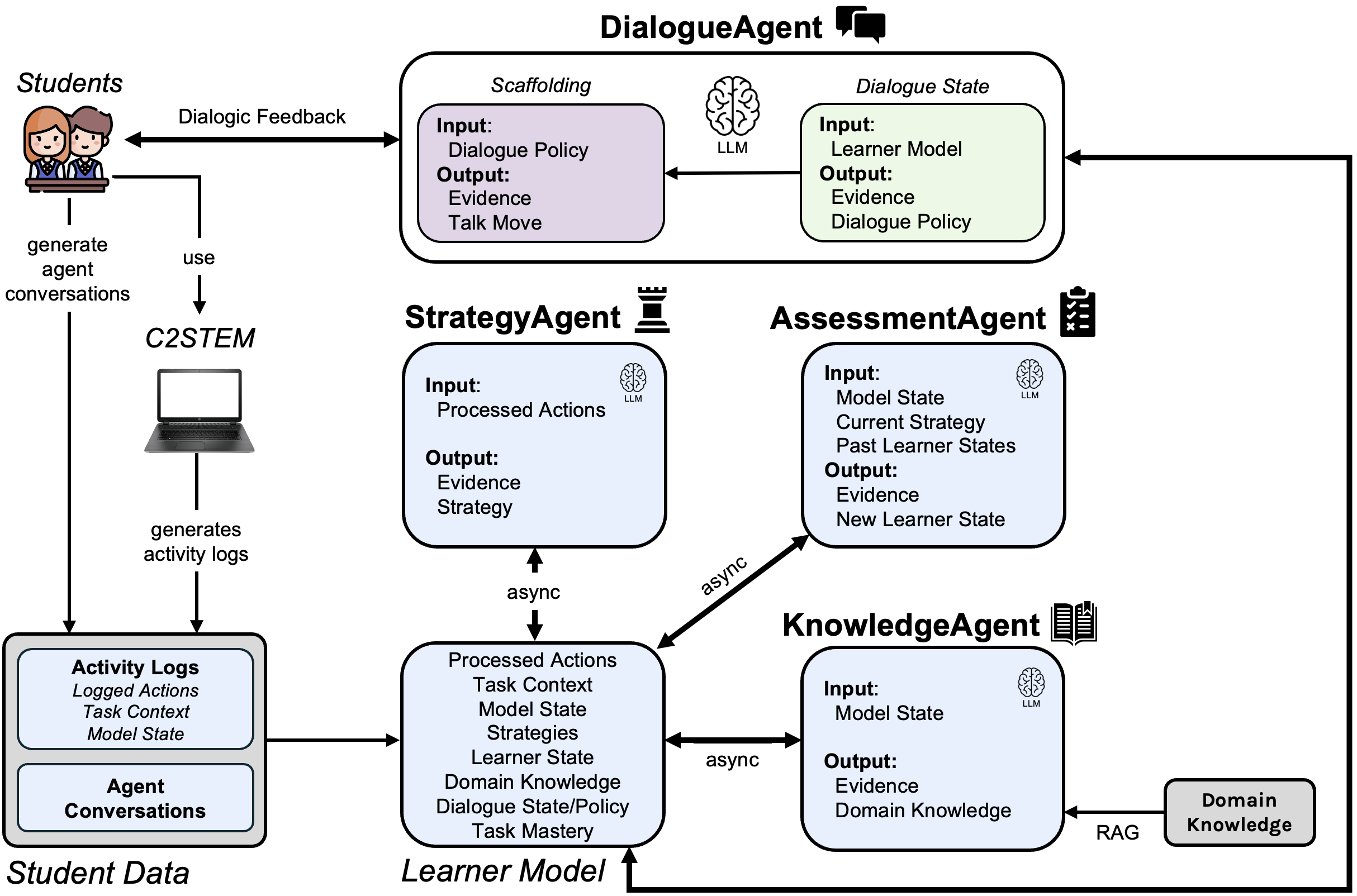} 
    \caption{Copa and its four sub-agents. Colors correspond to the EDF modules presented in Figure~\ref{fig:edf}. Gray boxes = data stores. Thick edges = two-way relationships.} 
    \label{fig:architecture} 
\end{figure}

Shown in Figure~\ref{fig:architecture}, Copa comprises four specialized sub-agents operating over a shared \emph{learner model} that is continuously updated through \emph{activity logs} and \emph{agent conversations}. This architecture directly maps onto the EDF modules via its color-coded sub-agents, with particular emphasis on the Evidence module (blue), which comprises three asynchronous agents that transform noisy log data into a coherent learner model. All four agents additionally emit \emph{evidence} with their outputs, using CoT to justify decisions and ensure interpretability.

The \textbf{StrategyAgent} analyzes temporal action patterns to infer cognitive problem-solving behaviors, distinguishing productive \emph{strategies} (e.g., \texttt{TINKERING}: systematic trial-and-error) from haphazard approaches (e.g., \texttt{DEPTH-FIRST ENACTING}: randomly building without periodic testing). The \textbf{AssessmentAgent} triangulates the current model state and strategy with past \emph{learner states} (e.g., \texttt{EXPLORING}, \texttt{DEBUGGING}) to generate an updated learner state. The \textbf{KnowledgeAgent} compares the model state to an expert reference to identify conceptual gaps and retrieves ZPD-aligned \emph{domain knowledge} (i.e., kinematics/computing information) via retrieval-augmented generation (RAG).

The \textbf{DialogueAgent} is the central orchestrator that integrates the Decision (green) and Feedback (purple) modules. These modules must operate synchronously, so housing them within a single agent minimizes latency, enabling asynchronous reasoning via the Evidence agents without interrupting conversational flow. Upon receiving a student query, the DialogueAgent identifies the students' \emph{dialogue state} (e.g., \texttt{DEMONSTRATES\_UNDERSTANDING}, where students verbalize conceptual knowledge to Copa). The agent then synthesizes the dialogue state and learner model to determine pedagogical intent via a \emph{dialogue policy}, e.g., \texttt{PROBE\_UNDERSTANDING} (i.e., students want a direct solution or demonstrate a misconception, prompting Copa to first probe their knowledge), \texttt{SUGGEST\_ACTION} (i.e., take a concrete action in C2STEM), or \texttt{PUSH\_LIMIT} (i.e., push students to extend and deepen their understanding). This policy is then used by the DialogueAgent to generate a \emph{talk move}, i.e., a response to students.
       
We evaluated LLMs from the Gemini, Claude, and GPT families and selected GPT-5 for optimal performance. Prompts were developed using Chain-of-Thought Prompting + Active Learning (CoTAL)~\cite{cohn2025cotal}, a human-in-the-loop prompt engineering procedure from prior work that uses chain-of-thought (CoT) reasoning, targeted few-shot exemplars, and iterative prompt refinement to link student input data to teacher and researcher instructions. Prompts were refined across four hour-long sessions with 10-15 researchers, who interacted with Copa using C2STEM and provided feedback that informed subsequent revisions.

All agents have specific theoretical motivations. For example, the KnowledgeAgent estimates students' ZPD and identifies the ``immediate next step'' needed based on task progress, while the DialogueAgent is instructed to ``promote self-efficacy through encouragement and praise'' (SCT) and ask ``probing questions'' (Social Constructivism). All task rubrics, as well as agent prompts, input features and their values, outputs, functionality, and theoretical foundations are detailed in our \href{https://github.com/claytoncohn/AIED26_Supplementary_Materials}{Supplementary Materials}\footnote{\url{https://github.com/claytoncohn/AIED26_Supplementary_Materials}}.

\section{Evaluation}
\label{sec:eval}

To evaluate Copa, we address the following Research Questions:

\begin{enumerate}
    \item Do Copa's scaffolds appropriately adapt as students increase task mastery?
    
    \item Does students' verbalized understanding align with their task mastery when interacting with Copa?

    \item Do students rely less on Copa as they achieve greater task mastery?
    
    \item Is Copa's feedback interpretable with respect to its evidentiary inferences and student input data?
\end{enumerate}

In collaboration with five STEM educators and cognitive scientists (hereafter, ``the educators''), we deployed Copa in a high school science classroom in Nashville (USA) with $n=33$ sophomore dyads\footnote{We conducted all analyses at the dyad level (one student-pair per computer).} (ages 15--16; 48\% male/female, 2\% no response; 68\% White, 28\% Hispanic, 20\% Black, 12\% Asian, 4\% Native American)\footnote{Several students identified as multiple ethnicities.} participating in the C2STEM kinematics curriculum over six weeks. Sessions met once weekly for two hours and began with instruction on 1- and 2-D kinematics---linking physics concepts (e.g., position, velocity, acceleration, $\Delta t$) with computational constructs (e.g., variable initialization and updating, loops, and conditionals)---followed by computational modeling tasks in C2STEM. Classes were taught by this paper's lead author with oversight from the educators. Our analyses focus on three tasks: a 1-D truck, a 2-D drone dropping one package, and a 2-D drone dropping two packages to different targets.

We collected 7,017 logged environment actions (mean $\approx 213$ per dyad), and 238 student-agent conversation turns (mean $\approx 7$ per dyad) across the three sessions, along with the CoT reasoning (i.e., ``evidence'' in Figure~\ref{fig:architecture}) associated with each agent decision. Students completed an anonymous exit survey. All findings were presented to three of the participating educators during a semi-structured Zoom discussion to solicit additional feedback. Together, these data constitute the primary dataset for our analyses, with all inferences manually triangulated through student conversations and screen recordings. Study participants provided informed consent/assent, student data were anonymized prior to analysis, and all procedures were approved by Vanderbilt University's IRB.

\subsection{RQ1: Scaffolding Adaptivity}

As students achieve greater task mastery, Copa's scaffolding must adapt by gradually fading in accordance with ZPD principles. At the outset, Copa should prioritize probing students' understanding to assess their current knowledge, approximate their ZPD, and address prior misconceptions. As students deepen their understanding, Copa should look ahead and offer concrete suggestions to help them extend their reasoning beyond their current level of competence.

To evaluate Copa's adaptivity (RQ1), we examined how its dialogue policies evolved within each session by analyzing the frequency with which policies were selected relative to quintiles of students' task mastery percentages (see Table~\ref{tab:rq1}). We used quintiles for more stable comparisons of agent behavior across meaningful phases of learning. Policy frequencies were normalized as a percentage of the total dialogue in each dyad's session and then averaged across mastery quintiles. For each dialogue policy, we computed Spearman's $\rho$ (due to the ordinal nature of the mastery data) to assess the strength of the relationship between shifts in dialogue policy and task mastery.

\begin{table}[t]
    \centering
    \caption{Dialogue policy adaptation across task mastery quintiles.}
    \begin{tabular}{lccc}
        \hline
        Dialogue Policy & Spearman's $\rho$ & Trend & $p$-value \\
        \hline
        \texttt{PROBE\_UNDERSTANDING}   & $-0.34$ & Decreasing & $0.034$ \\
        \texttt{SUGGEST\_ACTION}        & $0.33$  & Increasing & $0.039$ \\
        \texttt{PUSH\_LIMIT}            & $0.42$  & Increasing & $0.007$ \\
        \hline
    \end{tabular}
    \label{tab:rq1}
\end{table}

As shown in Table~\ref{tab:rq1}, Copa significantly ($\alpha = 0.05$) reduced its use of the \texttt{PROBE\_UNDERSTANDING} policy as mastery increased ($\rho = -0.34$), shifting toward implementation- and expansion-oriented support: \texttt{SUGGEST\_ACTION} ($\rho = 0.33$) and \texttt{PUSH\_LIMIT} ($\rho = 0.42$) increased in relative frequency. This demonstrates a pedagogical transition in which Copa moves from eliciting and assessing conceptual understanding to supporting its application through model construction as mastery increases---reflecting ZPD alignment as Copa shifts assistance toward challenges beyond learners' current competence and indicating \textbf{Copa's scaffolding appropriately adapted as task mastery increased}.

\subsection{RQ2: Understanding-Mastery Alignment}

Students' prioritization of task completion over conceptual understanding can present an overly optimistic picture of their learning, as students may appear to progress without being able to explain their code. From a social constructivist perspective, learning is evidenced not only by task performance but by learners' ability to articulate and negotiate meaning through dialogue and explanation. Evaluating whether students' task mastery is meaningfully related to their verbalized understanding while interacting with Copa is, therefore, paramount.

\refstepcounter{footnote}
\begin{figure}[htbp]
    \centering
    \includegraphics[width=1\linewidth]{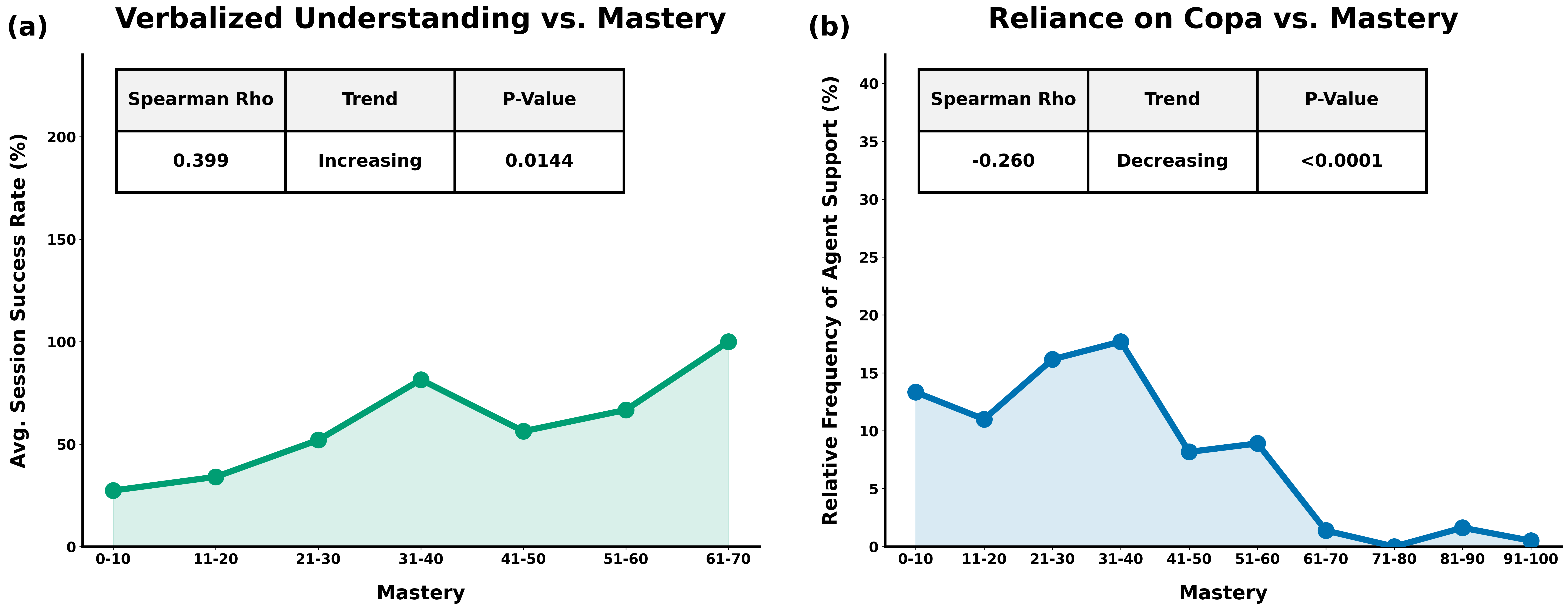}
    \caption{
        \textbf{(a)} Correlation between success rate\protect\footnotemark[\thefootnote] and mastery deciles (RQ2).\\
        \textbf{(b)} Correlation between student requests for agent support and mastery deciles (RQ3).}
    \label{fig:rqs_2_3}
\end{figure}
\footnotetext[\value{footnote}]{Copa chose not to probe students' knowledge if mastery was greater than 0.7.}

To address RQ2, we examined the relationship between students' response \emph{success rates}---defined as the proportion of instances students correctly explained their code via a \texttt{DEMONSTRATES\_UNDERSTANDING} dialogue state immediately following a \texttt{PROBE\_UNDERSTANDING} dialogue policy---and students' task mastery deciles. Deciles provide a finer-grained analysis of how students' verbally-demonstrated understanding varies across incremental levels of task mastery to assess alignment between the two. Data were normalized by averaging response success rate percentages for each dyad to mitigate interaction frequency bias. We again used Spearman's $\rho$ due to the ordinality of deciles.

Figure~\ref{fig:rqs_2_3}a shows the success rate increased significantly as mastery increased ($\rho = 0.40$), suggesting students were not merely ``gaming the system'': mastery gains were accompanied by an improved ability to verbally reason about their code and respond correctly to Copa, i.e., C2STEM progress was meaningfully aligned with conceptual understanding rather than superficial task optimization. Pursuant to Social Constructivism, articulating reasoning through dialogue reflects learners' active knowledge construction through interaction. Together, these findings demonstrate that \textbf{students' verbalized conceptual understanding aligned with their C2STEM task mastery}.

\subsection{RQ3: Student Reliance on Agent}

Students' control over when and how they interact with Copa promotes agency but risks overreliance. This can impede persistent learning and necessitates evaluating whether students use Copa in a manner consistent with productive support or continue to rely on the agent despite having developed sufficient mastery. As students progress, they should engage Copa less often, reflecting increasing independence and reduced reliance on external support. Pursuant to SCT, this transition reflects the development of self-regulation and self-efficacy skills.

To evaluate the appropriateness of students' reliance on Copa (RQ3), we examined how students' engagement with the agent changed across task mastery deciles. We computed the percentage of agent support (i.e., interactions) across deciles to quantify the proportion of Copa turns per session. To control for bias introduced by imbalanced interaction volumes, we normalized the data by averaging the agent-support proportion for each dyad across deciles. We again used Spearman's $\rho$ due to the ordinal nature of deciles.

Figure~\ref{fig:rqs_2_3}b shows students' requests for support were highest early on, with approximately 59\% of support occurring while mastery was low ($<40$\%). As mastery increased, we observed a significant drop in requests for agent feedback (Spearman's $\rho = -0.26$): \textbf{students increasingly relied on their own knowledge and collaboration with their partner rather than on agent support as mastery improved}, consistent with developing learner autonomy, self-regulation, and self-efficacy---central SCT tenets. 

\subsection{RQ4: Feedback Interpretability}

Agents must be transparent about how they make decisions in order for stakeholders to trust them; otherwise, they risk limited adoption. EDF's emphasis on interpretability aligns with ECD and Stealth Assessment, which require inferences about student knowledge to be grounded in observable evidence collected unobtrusively during interactions. Agent feedback must be traceable to the dialogue policy that generated it, the evidence identified by the agent, and ultimately to the raw student input data. To assess the interpretability of Copa's feedback (RQ4), we decompose Copa's internal reasoning chain (the agents' ``evidence'' outputs in Figure~\ref{fig:architecture}) into the three links between EDF modules:

\begin{equation*}
    \text{Student Input Data} \xrightarrow{\text{Link 1}} \text{Evidence} \xrightarrow{\text{Link 2}} \text{Decision} \xrightarrow{\text{Link 3}} \text{Feedback}
\end{equation*}

Link~1, referred to as \emph{Grounding}, measures the extent to which Copa's Evidence extraction reflects student input data. We evaluate Grounding using \emph{keyword recall}, defined as the percentage of semantically meaningful tokens from student log data that appear in the extracted evidence. We apply Porter stemming to account for morphological variation (e.g., accelerates $\approx$ acceleration), retaining only tokens longer than three characters with at least one alphabetic character. Significance is assessed via a permutation test ($n = 100$), comparing observed log-evidence pairs against randomly shuffled baselines.

Link~2, referred to as \emph{Alignment}, assesses semantic coherence between Copa's Decision (dialogue policy selection) and extracted Evidence. We compute semantic similarity using Sentence-BERT embeddings (all-MiniLM-L6-v2)---selected for its balance of size and performance---by encoding Copa's dialogue policy and dialogue state summaries, then comparing them via cosine similarity. Significance was evaluated via a permutation test ($n = 1000$)\footnote{Larger $n$ due to fewer utterances relative to environment actions in Link~1.}, comparing mean similarity of true evidence-policy pairs to shuffled baselines. Link~3, referred to as \emph{Faithfulness}, evaluates whether Copa's final Feedback adheres to its Decision. We assessed faithfulness by computing the SBERT-based cosine similarity between the dialogue policy and talk move, using the same Link~2 testing procedure.

\begin{table}[htb]
    \centering
    \caption{Interpretability results for all three EDF links.}
    \begin{tabular}{lcccr}
        \hline
        \textbf{Trace Component} & \textbf{Metric} & \textbf{Actual} & \textbf{Random} & \textbf{p-value} \\
        \hline
        Grounding (Data $\rightarrow$ Evidence) & Keyword Recall & \textbf{0.43} & 0.21 & $p < 0.001$ \\
        Alignment (Evidence $\rightarrow$ Decision) & SBERT Similarity & \textbf{0.64} & 0.39 & $p < 0.001$ \\
        Faithfulness (Decision $\rightarrow$ Feedback) & SBERT Similarity & \textbf{0.48} & 0.24 & $p < 0.001$ \\
        \hline
    \end{tabular}
    \label{tab:qta_results}
\end{table}

Table~\ref{tab:qta_results} shows that all three links exhibit statistically significant non-random structure ($p < 0.001$), with similarity scores for true pairs approximately doubling those of the baselines. The Grounding score indicates Copa's evidence extraction captured approximately twice as many meaningful tokens from the logs relative to the baseline (0.43 vs. 0.21). The Alignment score of 0.64 (vs. 0.39) indicates strong semantic coherence between the extracted evidence and the dialogue policy selections. The Faithfulness score of 0.48 (vs. 0.24) demonstrates that Copa's generated responses remain semantically aligned with the dialogue policy. Viewed through the lens of ECD and Stealth Assessment, these results indicate \textbf{Copa's decisions are interpretable with respect to how observable learner data is transformed into evidentiary representations, pedagogical decisions, and student-facing feedback.}

\subsection{Student and Teacher Perceptions}
    
Post-study survey data revealed students responded positively to Copa overall, agreeing that \textit{``the questions Copa asked were appropriate''} (mean $= 3.81$ on a five-point Likert scale) and Copa \textit{``asks questions that made me think about my model''} (mean $= 3.88$). This behavior was not always appreciated, as ratings for perceived agent understanding (\textit{``I feel like Copa understood what I was doing''}) and feedback utility (\textit{``I feel like Copa's suggestions were useful''}) were lower (mean $= 2.69$ and $2.77$, respectively). While students reported appreciating Copa's inquiry-based approach, they often became frustrated with the agent when their desire for direct answers was not met. This highlights a broader tension between principled support and students' agency and expectations.

Our findings aligned with educators' experiences, who agreed that (1) students become frustrated when they do not receive immediate answers (\textit{``I agree with your conclusion...students wanting immediate answers''})---an expectation shaped by students' prior ChatGPT experience, they noted; and (2) a persistent tension exists between instructional goals and student wishes, as students often \textit{``just [want]...the right answer...check the box and move on.''} They also remarked that students hold different expectations for agents than humans, underscoring the importance of students ``buying in'' to agent instructional intent. 

\section{Discussion and Conclusions}
\label{sec:discussion_conclusion}
In this paper, we presented EDF as a theory-driven adaptive scaffolding framework for LLM agents. Through Copa, we demonstrated EDF enables agent feedback that reflects students' evolving mastery while remaining interpretable, avoiding overreliance, and supporting knowledge verbalization in line with social-cognitive and social-constructivist principles. Copa's modular, agentic design provides a tight theory-to-architecture coupling, which we evaluated in an authentic classroom setting. Overall, we showed that (1) agentic, multi-agent LLM systems can be designed to operate in principled, interpretable, and pedagogically meaningful ways; and (2) EDF offers a viable framework for operationalizing learning theory within LLM-based agents toward these goals.

Theory emphasizes that effective support should help students build understanding rather than just providing answers. However, consistent with our prior work, students do not always value this~\cite{cohn2025personalizing}. Their frustration when denied direct answers reveals a design tension we term the \emph{helpfulness paradox}: pedagogical agents must be helpful enough to earn trust and invite use when needed, yet restrained enough to avoid encouraging cognitive offloading that harms durable learning. Effective LLM agents must therefore balance immediate utility with constraints that preserve independent problem solving.

From a methodological perspective, LLM pedagogical agent adaptation requires continuous scaffold adjustment based on learner evidence rather than fixed strategy execution. Recent work on agentic AI failures shows that agents often appear effective in testing but degrade in authentic use when they fail to adapt~\cite{jiang2025adaptation}. This further motivates evaluation in real classroom settings, where learners' shifting behaviors, needs, and goals expose limitations that controlled benchmarks often miss. Practically, however, real-world impact depends not only on system design but also on students' willingness to engage with agent support. As Thomas et al. (2025)~\cite{thomas2025llm} note, the value of agent feedback is ultimately constrained by whether students choose to use it. Fostering student buy-in and trust is therefore essential, as learning benefits are unlikely to emerge if students disengage from or resist agent feedback.

Our findings also point to a broader implication for evaluation: agents must be assessed across multiple dimensions. Prior agent evaluations often center on a single learner or agent attribute---e.g., theoretical alignment, learning gains, self-efficacy~\cite{cohn2026theory}---rather than evaluating agents holistically. In our study, adaptivity, understanding-mastery alignment, reliance, interpretability, and stakeholder perceptions are complementary, each capturing important, interconnected aspects. Accordingly, multi-lens agent evaluation must become the norm rather than the exception, as performance gains along a single dimension may obscure failures in other dimensions that are equally critical to meaningful learning.

The primary limitations of our work include evaluating scaffold adaptation solely as a function of task mastery rather than a broader set of learner attributes, and the absence of an RCT due to institutional constraints: reported relationships are correlational rather than causal. Our evaluation focuses on within-system behavior rather than comparison against alternatives, and although EDF is a general framework, we evaluate it through Copa only. Manual triangulation across log, screen, and video data suggests Copa supports self-regulation and produces pedagogically sound reasoning chains; however, rigorous stakeholder-centric evaluation across these axes remains future work.

This paper's central contribution is demonstrating that EDF can be used to operationalize LLM agents in pedagogically meaningful ways, as supported by our findings. In future work, we will address the limitations above by broadening evaluation to include self-regulation, self-efficacy, social dynamics, and in-depth stakeholder interpretations. Our findings and future research emphasize the importance of grounding LLM agents in learning theory, empirical evidence, stakeholder perspectives, and robust evaluation. We hope this work motivates further AIED research that moves beyond surface-level LLM interactions toward more adaptive and trustworthy agents that can support learning at scale.

\begin{credits}
\subsubsection{\ackname} 
We thank the Institute of Education Sciences (IES; award R305C240010) and National Science Foundation (NSF; awards IIS-2327708 and DRL-2112635) for their support. All opinions are those of the authors, not the IES or NSF.

\subsubsection{\discintname}
The authors have no competing interests to declare.

\end{credits}
%
%
%
\bibliographystyle{splncs04}
\bibliography{references}

\end{document}